\documentclass[a4paper,fleqn,usenatbib]{mnras}

\usepackage{newtxtext,newtxmath}
\usepackage[T1]{fontenc}
\usepackage{ae,aecompl}

\usepackage{graphicx}	% Including figure files
\usepackage{amsmath}	% Advanced maths commands
\usepackage{amssymb}	% Extra maths symbols
\usepackage{epstopdf}

\title[Evidence of Planet Around KIC 5095269]{Evidence for a planetary mass third body orbiting the binary star KIC 5095269}

\author[Getley et al.]{
A. K. Getley,$^{1}$
B. Carter,$^{1}$
R. King$^{1}$
and S. O'Toole$^{2}$
\\
$^{1}$Computational Engineering and Science Research Centre, University of Southern Queensland, Toowoomba Qld 4350 Australia \\
$^{2}$Australian Astronomical Observatory, PO Box 915, North Ryde NSW 1670 Australia
}

\date{Accepted XXX. Received YYY; in original form ZZZ}

\pubyear{2016}

\begin{document}
\label{firstpage}
\pagerange{\pageref{firstpage}--\pageref{lastpage}}
\maketitle

% Abstract of the paper
\begin{abstract}
In this paper, we report the evidence for a planetary mass body orbiting the close binary star KIC 5095269. This detection arose from a search for eclipse timing variations among the more than 2,000 eclipsing binaries observed by \textit{Kepler}. Light curve and periodic eclipse time variations have been analysed using \textit{Systemic} and a custom \textit{Binary Eclipse Timings} code based on the \textit{Transit Analysis Package} which indicates a $7.70\pm0.08M_{Jup}$ object orbiting every $237.7\pm0.1d$ around a $1.2M_\odot$ primary and $0.51M_\odot$ secondary in an 18.6d orbit. A dynamical integration over $10^7$ years suggests a stable orbital configuration. Radial velocity observations are recommended to confirm the properties of the binary star components and the planetary mass of the companion.
\end{abstract}

% Select between one and six entries from the list of approved keywords.
% Don't make up new ones.

%%%%%%%%%%%%%%%%%%%%%%%%%%%%%%%%%%%%%%%%%%%%%%%%%%

%%%%%%%%%%%%%%%%% BODY OF PAPER %%%%%%%%%%%%%%%%%%
\section{Introduction}

Planet formation is widely considered to be dominated by core accretion, but an alterative disc instability mechanism has been proposed to explain gas giant planets and brown dwarfs \citep{boss2012, chabrier2014}. In addition, while it is convenient to set a mass divide (at around 13 Jupiter masses) between planets and deuterium burning brown dwarfs \citep{burgasser2008}, it has been argued that a separation of planets and brown dwarfs based on the formation mechanism is more physically meaningful \citep{nordlund2011}. A key way to constrain planet formation models is to test their predictions as to the frequency, masses, orbits, and stability of planets orbiting eclipsing binary stars, whose mutual eclipses can also provide accurate host star properties. The problem to be solved however is to find such planets, as to date very few such circumbinary planets have been found \citep{sigurdsson2003, correia2005, lee2009, doyle2011, kostov2016}. In particular the discovery of the long period transiting circumbinary planet Kepler-1647b may represent an example of a large population of distantly orbiting massive planets orbiting close binary stars \citep{kostov2016}. Our research therefore represents the initial results of a search for eclipsing binary planets that uses eclipse timings to enable planets orbiting above or below the stellar orbital plane to be detected.

Contained in the `Kepler Eclipsing Binary Catalog' are more than 2,000 eclipsing binaries that have been observed over the life of the KEPLER mission \citep{prsa2011, slawson2011}. Predominately detached binaries stars, i.e. binary stars with a morphology classification of less that 0.5, account for almost half of the systems in the eclipsing binary catalog. The high precision observations that were performed allows an eclipse time study to be performed. Eclipsing binary stars that are detached and isolated should have eclipses that occur a constant and predictable time apart. Plotting the observed eclipse time (O) minus the calculated eclipse time (C) against a best-fitting linear ephemeris, variations from this constant time may be able to be seen. Periodic variations may be the result of a third body orbiting the binary \citep{beuermann2010}.

In systems that show periodic variations, the properties of the binary stars need to be estimated in order to fit and determine the characteristics of any additional bodies. Estimates for the masses of the binary stars are calculated from the colour data given in the Kepler data and modelling the light curve in JKTEBOP \citep{southworth2004}. Colours and masses for spectral types are given in \cite{pecaut2013}. With a mass ratio and mass estimate from the system colours, individual masses can be worked out. By using Systemic \citep{meschiari2009, meschiari2010} a system can be set up with the masses and characteristics of the binary stars. From here additional bodies can be added and fit to determine if characteristics can account for eclipse time variations.

In this paper, we report on the results of an eclipse time study of a specific Kepler system, KIC 5095269, and the follow up Systemic study in order to determine the characteristics of a third body. We propose the existence of a third body around KIC 5095269 with a mass of $7.70\pm0.08$ Jupiter masses.

\section{O-C Production and Identification}

We used the Kepler data to produce O-C diagrams to study eclipse timing variations. Detached eclipsing binaries were selected in order to minimise variations from within the system itself. A primary eclipse occurs when the larger star passes in front of the smaller star, while a secondary eclipse occurs when the smaller star passes in front of larger star. The time of as many primary eclipses and secondary eclipses as possible must be determined in order to perform an eclipse time variation study. We created a program, called BET or Binary Eclipse Timings, to determine eclipse times. BET is based on the software Transit Analysis Package or TAP \citep{gazak2012} and uses the analytic formulae for the transit or eclipse of a star which are found in \cite{mandel2002}. The analytic formulae in \cite{mandel2002} describe a system of two objects during various points in its orbit. The objects can be a star and a planet (i.e. describing transits) or two stars (i.e. describing eclipses). The systems are described using the parameters: orbital period, the radius ratio of the two objects, scaled semi-major axis, orbital inclination, orbital eccentricity, argument of periastron, mid-time of eclipse/transit and two parameters specifying quadratic limb darkening. BET detects eclipses from the Kepler data and uses the analytic formulae to accurately determine the mid-eclipse times of a system. 

With the observed eclipse times of a system determined, calculated eclipse times are needed in order to produce an O-C diagram. Since the time between eclipses should be constant, a calculated eclipse time can be found with the equation:

\begin{equation}
    T_{n} = P \times n + T_{0}
    \label{eq:linearephem}
\end{equation}

Where P is the period of the system, n is the cycle number and T\textsubscript{0} is the initial eclipse time.

Equation ~(\ref{eq:linearephem}) can be modified to account for primary and secondary eclipses and take the form seen in ~(\ref{eq:preciseephem}).

\begin{equation}
    \begin{split}
        T_{np} = P \times n + T_{0p},
        \\
        T_{ns} = P \times n + T_{0s}.
        \label{eq:preciseephem}
    \end{split}
\end{equation}

Where T\textsubscript{0p} is the initial primary eclipse, T\textsubscript{0s} is the initial secondary eclipse, n is the cycle number and P is the period of the system which is common to both primary and secondary eclipses. 

By performing a least-squares best-fit to the observed eclipse times with ~(\ref{eq:preciseephem}) the best-fit period and initial eclipse times will be found. Expected eclipse times can then be calculated. By plotting the observed eclipse time minus the calculated eclipse time against the predicted eclipse time, variations from the expected may be observed. These variations have been separated into five different, custom defined, categories based on their O-C diagrams: No or Irregular variations, Periodic variations, Sudden period flips, Long term trends and out of phase long term trends. Variations may be caused by star spots \citep{orosz2012}, apsidal motion \citep{beuermann2010} or dynamical interactions \citep{borkovits2003}. It is also possible that periodic variations are caused by the effects of a third body \citep{beuermann2010}.

The times of observations in the \textit{Kepler} data is in Barycentric Julian Date (BJD) which is the Julian Date that has been corrected for the effects of the Earth's orbit. This correction will prevent Earth's orbit from appearing in the O-C diagrams. During the eclipse timing study, systems with no or irregular variations could be seen. These O-C variations would range from 0 to approximately 30 seconds and appear with no recurring pattern. Systems that have O-C variations larger than 30 seconds and particularly those that exhibit periodic O-C variations that are suspected to be caused by the addition of a third body should be prioritised for futher investigation. However as apsidal motion may also be the cause of periodic variation \citep{beuermann2010} it can't be assumed that third bodies are the cause of the O-C variations. In the hunt for planets, small amplitude variations (i.e. a few minutes) are also prioritised over larger variations as larger objects (i.e. stars) will have more of an effect on binary stars than smaller objects (i.e. dwarf stars or planets). 

\section{Following Up on Identified O-C Diagrams}

With an O-C diagram showing periodic variability, the next task is to try to determine the cause of the variability. In this study the software Systemic \citep{meschiari2009, meschiari2010} has been used to model the system and estimate the characteristics of a third based on its effect on the binary stars. Systemic can be used to model eclipse and transit timing variations. In order to accurately determine the properties of any potential third bodies the mass for the primary and secondary stars must be estimated. Radial velocity data is needed to determine the mass of binary stars however these data can be difficult to obtain for Kepler stars without the use of a large telescope. As a result mass estimates for the binary stars were calculated based on the data in the Kepler database and the light curve of the system. 

JKTEBOP \citep{southworth2004} was used to find the best fit to model the light curve in order to determine/estimate the parameters of the system such as the orbital period, mass ratio of the binary stars and inclination of the system. Other data for the binary star systems such as the V-K colour can be used to help validate and guide the mass estimates of the binary star. By using the colours of the system and the mass ratio and other property estimates from JKTEBOP a system can be set up in Systemic to determine the properties of a potential third body. If the colours fall between two star types, the larger masses can be used and as a result the mass of any third body present should be an upper estimation of the mass. JKTEBOP was selected as it is capable of fitting the parameters of a system, including limb darkening and mass ratio.

\section{Results}

\begin{figure*}
    \includegraphics[width=\columnwidth]{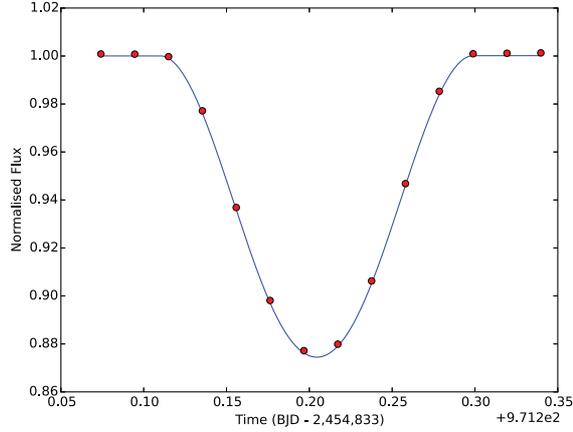}
    \caption{Example of a primary eclipse fit by BET (blue solid line) to the data obtained from KEPLER (red circles).}
    \label{fig:beteclipse}
\end{figure*}

Having processed the detached eclipsing binary stars from the Kepler `Eclipsing Binary Catalog', the O-C diagrams were then classified. One of the systems identified from the O-C diagrams as a potential host to a third body was KIC 5095269. The primary eclipse times and the errors reported by the fitting function from BET can be seen in Table ~\ref{tab:eclipsetimes} and an example eclipse fit from BET is shown in Fig. ~\ref{fig:beteclipse}. Secondary eclipses were too shallow and unable to be fit and have accurate times determined.

\begin{table*}
    \centering
    \caption{Table of eclipse times for KIC 5095269.The eclipse time is in BJD - 2,454,833. The eclipse times are used to produce an O-C diagram to look for eclipse timing variations. A sample of the table is shown here, the full table is available online.}
    \label{tab:eclipsetimes}
    \begin{tabular}{lc}
    \hline
    Eclipse Time & Error\\
    (BJD - 2,454,833) & (Days)\\
    \hline
    133.865203 & $2.78 \times 10^{-5}$\\
    152.478622 & $2.76 \times 10^{-5}$\\
    171.091481 & $2.72 \times 10^{-5}$\\
    189.702749 & $2.71 \times 10^{-5}$\\
    208.313380 & $2.67 \times 10^{-5}$\\
    226.924859 & $2.71 \times 10^{-5}$\\
    245.537254 & $2.63 \times 10^{-5}$\\
    264.150062 & $3.12 \times 10^{-5}$\\
    \hline
    \end{tabular}
\end{table*}

\begin{figure*}
    \includegraphics[width=\columnwidth]{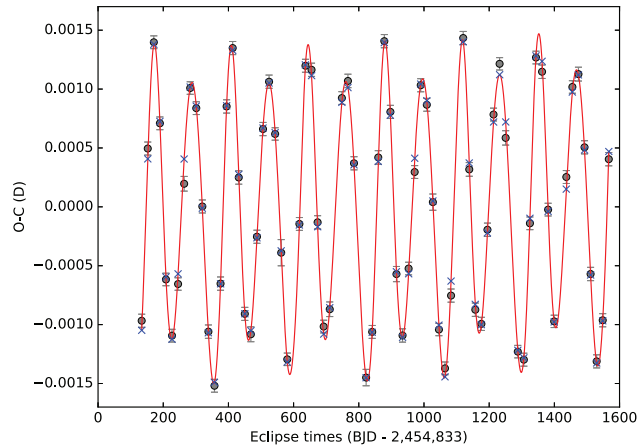}
    \caption{The O-C points and best-fit model to explain the eclipse timing variations. The grey circles are the O-C points for KIC 5095269 while the red line is the modelled eclipse time variations. A blue X marks the best-fit O-C points. Systemic takes the entered eclipse times, integrates the system to find the eclipse time closest to the entered time and plots the O-C value. Non-Keplerian dynamics and the orbital characteristics of the third body (visualised in Fig. ~\ref{fig:visualisation}) are thought to be the source of the variation in the maxima and minima in this O-C diagram.}
    \label{fig:systemicfit}
\end{figure*}

Using the data found in Table ~\ref{tab:eclipsetimes} and fitting using the functions found in ~(\ref{eq:preciseephem}) the period of the system, P, was found to be 18.611957 days and the initial primary eclipse, T\textsubscript{0p}, occurred at 133.866170 (BJD - 2,454,833). The secondary eclipses were found to be too shallow to accurately fit and as such no secondary eclipse times were obtained. The O-C diagram and data for the primary eclipses in KIC 5095269 is shown by the circles  in Fig. ~\ref{fig:systemicfit}. The O-C diagram shows periodic variability that has a period of approximately 120 days with variations in the eclipse times of up to approximately 2 minutes.

\begin{figure*}
    \includegraphics[width=\columnwidth]{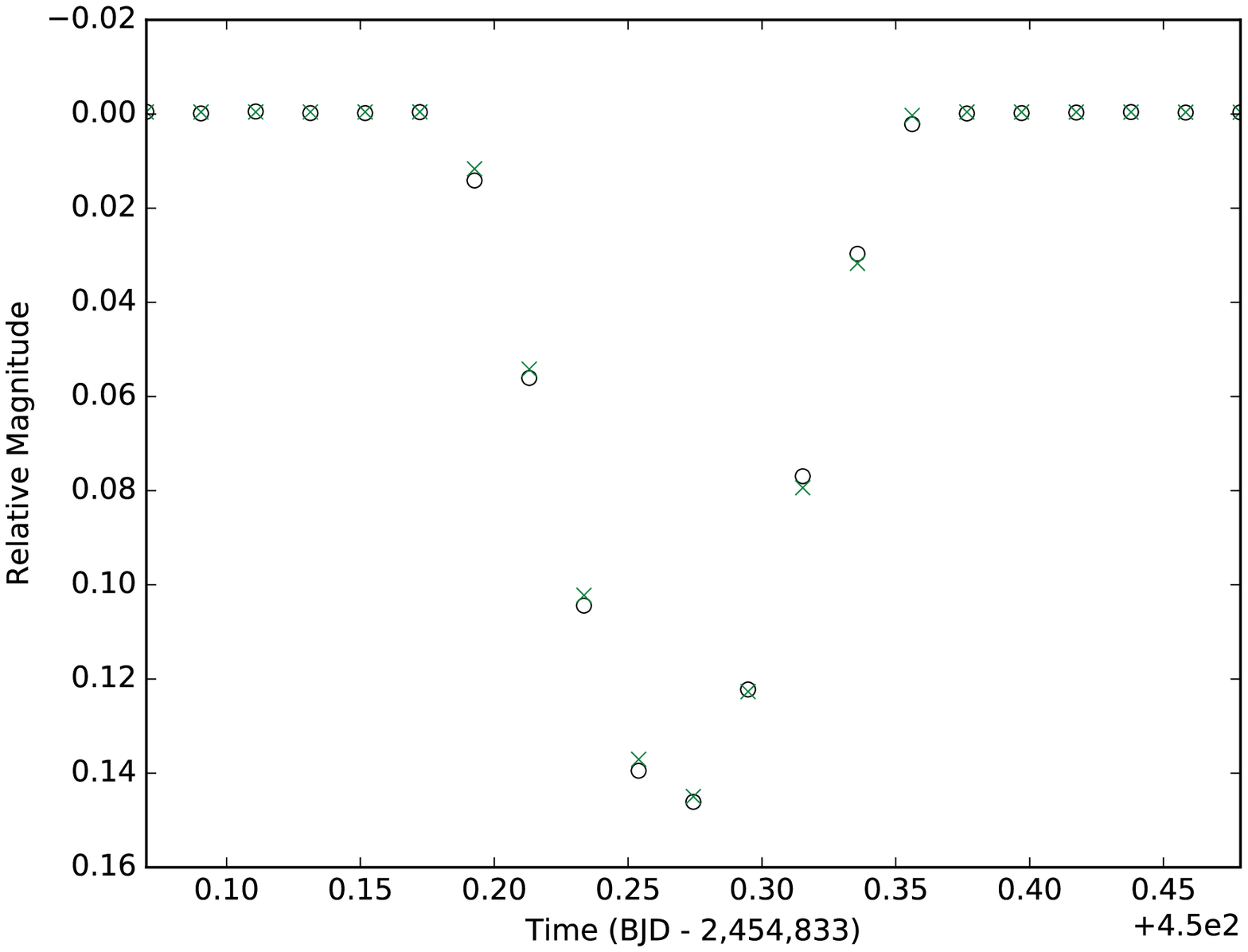}
    \caption{A small section of the light curve from KEPLER and the model from JKTEBOP showing an eclipse. The light curve data from KEPLER is shown by the circles, while the model data obtained from JKTEBOP is shown by x.}
    \label{fig:jktebop}
\end{figure*}

In addition to the orbital period of 18.611957 days and an initial eclipse occuring at 133.866170 (BJD - 2,454,833), modelling the light curve in JKTEBOP \citep{southworth2004} found a mass ratio of approximately 0.421 and an inclination of 80.02 degrees. An eclipse from the KEPLER data with the modelled light curve from JKTEBOP can be seen in Fig. ~\ref{fig:jktebop}. With a KEPLER magnitude of 13.528 and a 2MASS K magnitude of 12.215 this system has a V - K value of 1.313. This V - K value approximately matches an F7 star \citep{bessell1988}. If the secondary star were much hotter than an M star it would have an impact on the K magnitude of the system, and therefore the V - K value. This is not consistent with observations. With an F7V primary star with a mass of 1.21$M_{\odot}$ \cite{pecaut2013} and a mass ratio of 0.421, the secondary star would have a mass of 0.51$M_{\odot}$. This is consistent with the mass of an M star as the V - K colour suggests.

Systemic \citep{meschiari2009, meschiari2010} was then used to set up a representative system with the main (larger) star being set to a mass of 1.21 Solar Masses. `Planet 1' was set up with the characteristics of the secondary star and the system i.e. a mass of 0.51 Solar Masses, an orbital period of 18.611957 days and an inclination of 80.0235 degrees. The masses as well as the orbital period and inclination of the binary stars were fixed while all the other parameters were free to be fit by the program. An additional planet (`Planet 2') was added to the system with the period of the O-C variability set as the period of the planet. All parameters for the additional body were also free to be fit. The eclipse times were loaded into Systemic and the uncertainties in the eclipse times were doubled in Systemic in order to estimate the true uncertainty in the eclipse times. A best-fitting was then performed by Systemic to find the values for the system that best explains the O-C variability.

\begin{figure*}
    \includegraphics[width=\columnwidth]{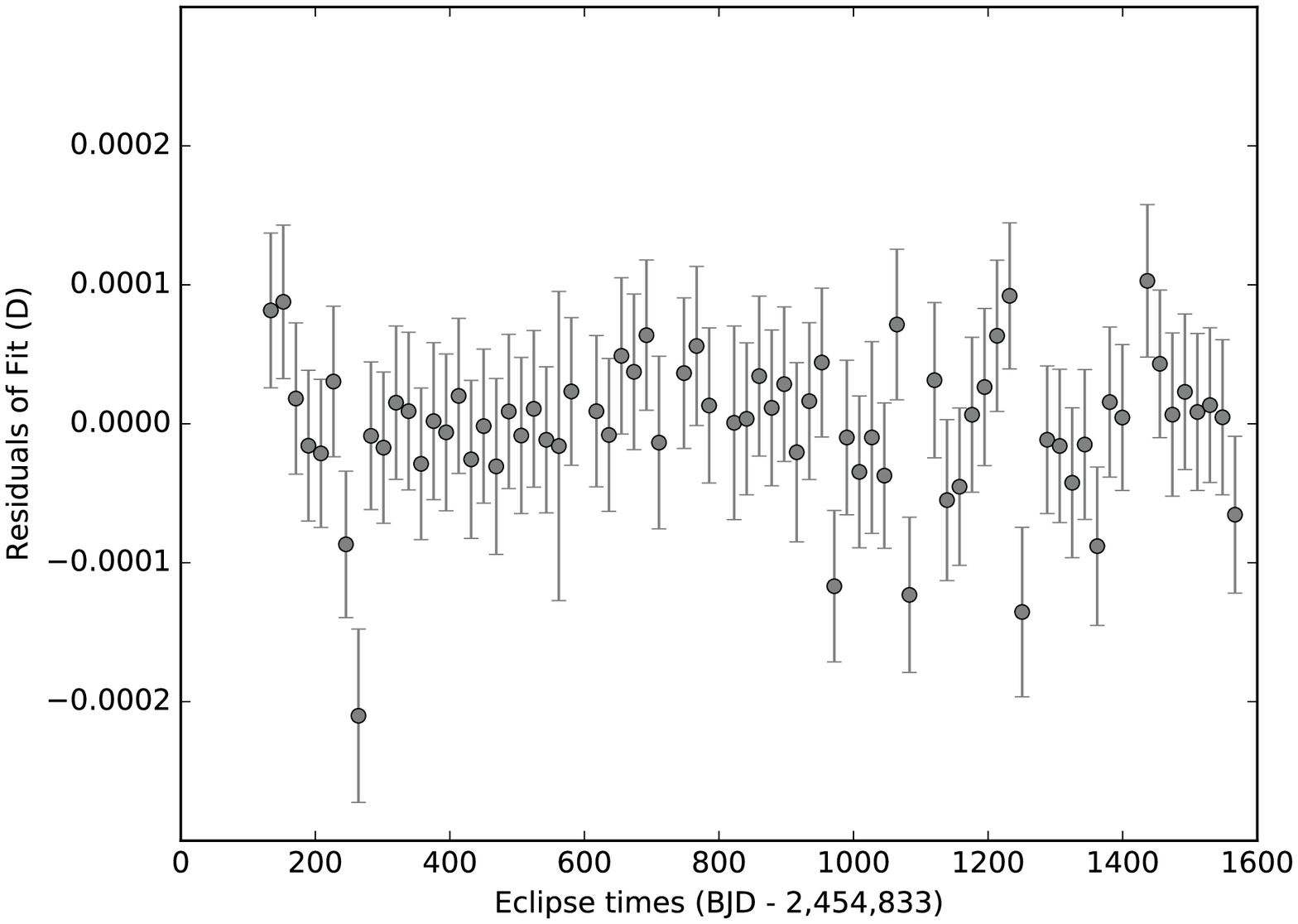}
    \caption{Residuals of the Systemic fit showing the difference between the observed eclipse times and the modelled eclipse times. The smaller the values of the residuals, the closer the modelled eclipse times are to the observed eclipse times.}
    \label{fig:residuals}
\end{figure*}

The results of the Systemic best-fitting can be seen in Fig. ~\ref{fig:systemicfit}, the residuals of the fit are shown in Fig. ~\ref{fig:residuals} and the data for the binary system is shown in Table ~\ref{tab:fitdatabinary} while the data for the third body is shown in Table ~\ref{tab:fitdataplanet}. The properties of the binary system were entered into PHysics Of Eclipsing BinariEs or PHOEBE \citep{degroote2013, prsa2005} to view a synthetic light curve. PHOEBE suggested an inclination of approximately 86.5 degrees was required to view a primary eclipse but no/minimal secondary eclipse as can be seen in the light curve. The inclination of the system was kept at the 80.02 degrees suggested by JKTEBOP as this was found by fitting the light curve however the larger inclination from PHOEBE was noted.

\begin{table*}
    \centering
    \caption{Table of data provided by the Systemic fit for the binary system. The mass of the primary and secondary stars, the orbital period of the binary stars and the inclination were fixed at pre-calculated values while the rest of the parameters were free to be fit. The Median and Median Absolute Deviation (MAD) values were also determined by Systemic.}
    \label{tab:fitdatabinary}
    \begin{tabular}{lccr}
    \hline
    Property & Best-Fit Value & Median Value & MAD Value\\
    \hline
    Primary Star Mass ($M_{\odot}$) & 1.21 & N/A & N/A\\
    Secondary Star Mass ($M_{\odot}$) & 0.51 & 0.51 & $1.24 \times 10^{-3}$\\
    Orbital Period (d) & 18.61196 & 18.61196 & $1.55 \times 10^{-7}$\\
    Mean anomaly (deg) & 7.44 & 7.44 & $1.31 \times 10^{-4}$\\
    Eccentricity & 0.246 & 0.246 & $9.3 \times 10^{-6}$\\
    Long. Peri (deg) & 22.82 & 22.82 & $2.23 \times 10^{-4}$\\
    Inclination (deg) & 80.0 & 80.0 & 0.02\\
    Node (deg) & 305.54 & 305.54 & $2.3 \times 10^{-4}$\\
    \hline
    \end{tabular}
\end{table*}

\begin{table*}
    \centering
    \caption{Table of data provided by the Systemic fit for the third body. All of the parameters were free to be fit. The Median and Median Absolute Deviation (MAD) values were also determined by Systemic.}
    \label{tab:fitdataplanet}
    \begin{tabular}{lccr}
    \hline
    Property & Best-Fit Value & Median Value & MAD Value\\
    \hline
    Mass ($M_j$) & 7.698 & 7.693 & 0.054\\
    Orbital Period (d) & 237.70817 & 237.68977 & 0.08237\\
    Mean anomaly (deg) & 290.92 & 289.44 & 2.34\\
    Eccentricity & 0.0604 & 0.0603 & 0.0021\\
    Long. Peri (deg) & 27.67 & 29.03 & 2.07\\
    Inclination (deg) & 105.92 & 105.83 & 0.98\\
    Node (deg) & 64.19 & 64.10 & 0.28\\
    \hline
    \end{tabular}
\end{table*}

\begin{figure*}
    \includegraphics[]{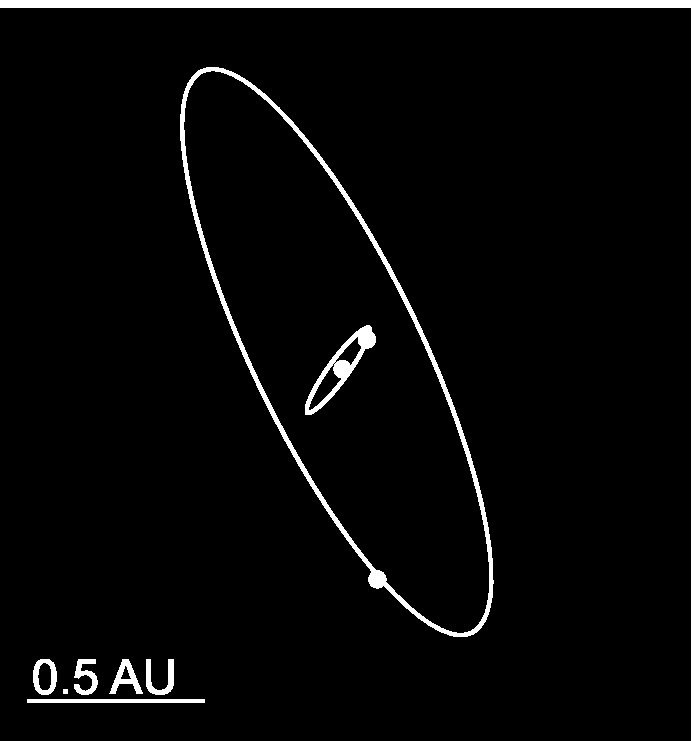}
    \caption{A visualisation of the system as determined by the program Systemic. The primary and secondary stars are at the centre and the third body is found to be orbiting both stars.}
    \label{fig:visualisation}
\end{figure*}

\section{Discussion}

The results of the Systemic fit indicate that a third body with an orbital period of $237.70817\pm0.12213$ days and a mass of $7.70\pm0.08$ Jupiter masses could account for the eclipse timing variations seen (all errors are quoted to a single standard deviation unless otherwise noted). The mass of this third body is expected to be an upper limit and suggests the third body is actually a planet orbiting the binary stars. The best fit orbit of the binary stars is 18.61196 days. The eccentricity of the binary stars and planet as determined by Systemic was found to be 0.246 and $0.060\pm0.003$ respectively. The data produced from Systemic found the best fit to match the eclipse timing variations occurs when there is a third body with a mass of $7.70\pm0.08$ Jupiter masses. This mass is below the proposed planet/brown dwarf boundary of roughly 13 Jupiter masses \citep{burgasser2008} and as such the third body is regarded as a planet rather than a brown dwarf. Though how this object formed may determine whether it is a large planet or a small brown dwarf \citep{nordlund2011}. If this third body is confirmed to be a circumbinary planet it would join a small number of previously confirmed circumbinary planets \citep{sigurdsson2003, correia2005, lee2009, doyle2011, kostov2016}, and this body would have one of the largest masses of these circumbinary planets.

With the period of the O-C variability being approximately 120 days it was expected that the orbital period of the third body would be approximately the same. However the Systemic fit was significantly better (both visually and by reduced $\chi^2$ value) with a period of 237.7 days and therefore this fit was chosen as the optimal fit. Systemic incorporates non-keplerian dynamics found in \cite{fabrycky2010} and it is thought that the orbital characteristics of the planet (visualised in Fig. ~\ref{fig:visualisation}) coupled with non-keplerian dynamics are the reason the orbital period of the planet varies from what was expected and also accounts for the variation seen in the maxima and minima of the O-C diagram in Fig. ~\ref{fig:systemicfit}. The orbital period of the planet approximately matches the period of the variation in the maxima and minima of the O-C diagram.

The found system was tested in order to check the robustness of the fit. By testing the binary star system in PHOEBE we can confirm that the values for the binary star system are reasonable. The inclination of the system was changed in Systemic to 86.5 degrees as found by PHOEBE. The results of the third body remain consistent with the mass of the third body changing to 7.72 Jupiter masses. The next test involved changing the mass of the primary and secondary stars to determine if the best-fit system correspondingly changed the mass of the third body. The mass ratio between the stars and the planet was found to remain the same regardless of the actual masses used. During the robustness test, the mass ratio of the stars and planet was also changed. In all cases a third body with significantly less mass than the binary stars was able to account for the eclipse timing variations seen. With a mass ratio between the binary stars of 0.7889 (i.e. the primary remains at our estimated value and the secondary star mass set to the largest value allowed by Systemic for additional bodies of 1000 Jupiter masses) the best fit mass for the third body was found to be 10.35 Jupiter masses which is still below the putative planet/brown dwarf boundary of 13 Jupiter masses. With a mass ratio 0.01 (i.e the secondary star has a mass that is 0.01 times the mass of the primary star) the best fit mass for the third body is 5.04 Jupiter masses. In all of the tests performed, with varying properties of the binary system, a planetary mass third body is capable of producing the eclipse timing variations seen. The effect of limb darkening on the mass ratio of the binary stars was also analysed with JKTEBOP. It was found that as the amount of limb darkening of both stars increased, the mass ratio decreased. It was also found that increasing the amount of limb darkening on the primary star only, resulted in a lower mass ratio, while increasing the amount of limb darkening on the secondary star only, had a very minor effect and slightly increased the mass ratio.

\begin{figure*}
    \includegraphics[width=\columnwidth]{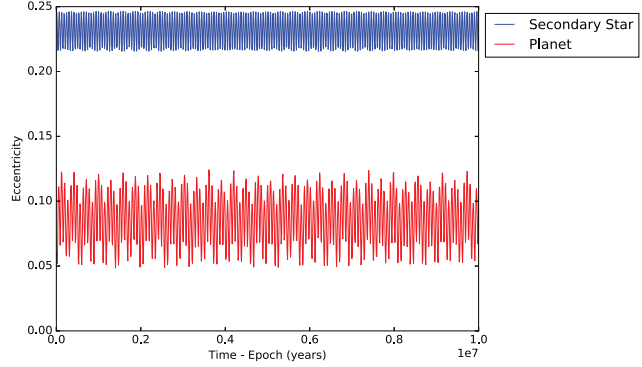}
    \caption{Eccentricity of the secondary star and planet after integration of the Systemic system over a period of $10^7$ years. The secondary star has very little variation in eccentricity. The planet has some variation in the eccentricity.}
    \label{fig:ecc}
\end{figure*}

\begin{figure*}
    \includegraphics[width=\columnwidth]{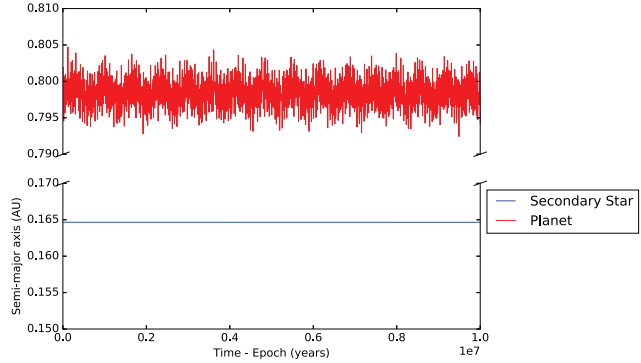}
    \caption{Semi-major axis of the secondary star and planet after integration of the Systemic system over a period of $10^7$ years. The secondary star has an almost constant semi-major axis while the planet has some variation in the semi-major axis.}
    \label{fig:sma}
\end{figure*}

The stability of the proposed orbits are important in determining whether the proposed orbits are the correct interpretation of the eclipse timing variations \citep{hinse2014}. The system was integrated over a period of $10^7$ years to determine the long term stability of the system. Systemic \citep{meschiari2009, meschiari2010} was used to perform the integration on the best-fit system found. The eccentricity results of the integration can be found in Fig. ~\ref{fig:ecc} while the semi-major axis results of the integration can be found in Fig. ~\ref{fig:sma}. The semi-major axis of the planet was found to vary between 0.795AU and 0.805AU while the eccentricity was found to fluctuate between 0.05 and 0.13 and indicates the planet orbits the binary stars in an almost circular orbit. As the orbits were found to be stable over large time periods, a planet in the proposed configuration is unlikely to be ejected from the system and is therefore the likely source of the eclipse timing variations seen \citep{hinse2014}.

With a proposed inclination of $105.92\pm1.45^{\circ}$ the probability of transits occurring needs to be considered. \cite{kane2008} presented an analysis of the effect of orbital parameters, specifically eccentricity and argument of periastron, on the probability of a transit as a function of the orbital period. The probability of a transit occurring for a planet in a circular orbit drops dramatically as the orbital period of the planet increases. As the planet found in this system has a proposed orbit of approximately 237 days and a nearly circular orbit the expected transit probability is less than approximately 0.01. By viewing the light curve no transits can be seen to occur which can be expected with such a low probability of a planet in this orbit transiting the parent stars.

\section{Conclusions}

In this paper, we presented the evidence for a third body below whose mass the planet/brown dwarf boundary \citep{burgasser2008} around KIC 5095269. An eclipse timing variation study was performed on the KEPLER detached binaries where KIC 5095269 was found to exhibit periodic eclipse time variations. As eclipse time variations may be the result of a third body \citep{beuermann2010} estimates for the mass of the binary stars were calculated and a model was produced with Systemic \citep{meschiari2009, meschiari2010}.

The model produced by Systemic suggests the cause of the eclipse timing variations is a third body with a mass of $7.70\pm0.08$ Jupiter masses. Based on the proposed planet/brown dwarf mass boundary \citep{burgasser2008} we propose that this third body is a planetary candidate however how this object formed may determine whether it is a planet or a brown dwarf \citep{nordlund2011}. The system was also found by Systemic to be stable over a period of $10^7$ years with an eccentricity of the third body that varies between 0.05 and 0.13. No transits could be seen to occur within the light curve although a planet with an orbital period of approximately 237 days has a probability of transit of less than 0.01.

 In the future, we will be attempting to obtain Radial Velocity data for the system in order to further constrain the properties of the system. This planetary candidate has provided us with a template, both the features of O-C variations and the method for analysis, to use for future systems with periodic variations. Systems with low mass secondary stars, combined with small O-C variations are the best chance for detecting planetary sized bodies via the eclipse time variation method.

\section*{Acknowledgements}
\addcontentsline{toc}{section}{Acknowledgements}

We would like to thank the referee and editor for their feedback and for helping us to fine tune the details of the paper. We would also like to thank Stefano Meschiari for his assistance with Systemic.

%%%%%%%%%%%%%%%%%%%%%%%%%%%%%%%%%%%%%%%%%%%%%%%%%%

%%%%%%%%%%%%%%%%%%%% REFERENCES %%%%%%%%%%%%%%%%%%

\bibliographystyle{mnras}
\bibliography{getley2016}

\begin{thebibliography}{}
\makeatletter
\relax
\def\mn@urlcharsother{\let\do\@makeother \do\$\do\&\do\#\do\^\do\_\do\%\do\~}
\def\mn@doi{\begingroup\mn@urlcharsother \@ifnextchar [ {\mn@doi@}
  {\mn@doi@[]}}
\def\mn@doi@[#1]#2{\def\@tempa{#1}\ifx\@tempa\@empty \href
  {http://dx.doi.org/#2} {doi:#2}\else \href {http://dx.doi.org/#2} {#1}\fi
  \endgroup}
\def\mn@eprint#1#2{\mn@eprint@#1:#2::\@nil}
\def\mn@eprint@arXiv#1{\href {http://arxiv.org/abs/#1} {{\tt arXiv:#1}}}
\def\mn@eprint@dblp#1{\href {http://dblp.uni-trier.de/rec/bibtex/#1.xml}
  {dblp:#1}}
\def\mn@eprint@#1:#2:#3:#4\@nil{\def\@tempa {#1}\def\@tempb {#2}\def\@tempc
  {#3}\ifx \@tempc \@empty \let \@tempc \@tempb \let \@tempb \@tempa \fi \ifx
  \@tempb \@empty \def\@tempb {arXiv}\fi \@ifundefined
  {mn@eprint@\@tempb}{\@tempb:\@tempc}{\expandafter \expandafter \csname
  mn@eprint@\@tempb\endcsname \expandafter{\@tempc}}}

\bibitem[\protect\citeauthoryear{{Bessell} \& {Brett}}{{Bessell} \&
  {Brett}}{1988}]{bessell1988}
{Bessell} M.~S.,  {Brett} J.~M.,  1988, \mn@doi [PASP] {10.1086/132281}, \href
  {http://adsabs.harvard.edu/abs/1988PASP..100.1134B} {100, 1134}

\bibitem[\protect\citeauthoryear{{Beuermann, K.} et~al.,}{{Beuermann, K.}
  et~al.}{2010}]{beuermann2010}
{Beuermann, K.} et~al., 2010, \mn@doi [A&A] {10.1051/0004-6361/201015472}, 521,
  L60

\bibitem[\protect\citeauthoryear{{Borkovits, T.}, {Erdi, B.}, {-Dajka, E.
  Forgac}  \& {Kovacs, T.}}{{Borkovits, T.} et~al.}{2003}]{borkovits2003}
{Borkovits, T.} {Erdi, B.} {-Dajka, E. Forgac}  {Kovacs, T.} 2003, \mn@doi
  [A&A] {10.1051/0004-6361:20021688}, 398, 1091

\bibitem[\protect\citeauthoryear{Boss}{Boss}{2012}]{boss2012}
Boss A.~P.,  2012, \mn@doi [MNRAS] {10.1111/j.1365-2966.2011.19858.x}, 419,
  1930

\bibitem[\protect\citeauthoryear{{Burgasser}}{{Burgasser}}{2008}]{burgasser2008}
{Burgasser} A.~J.,  2008, \mn@doi [Phys. Today] {10.1063/1.2947658}, \href
  {http://adsabs.harvard.edu/abs/2008PhT....61f..70B} {61, 70}

\bibitem[\protect\citeauthoryear{{Chabrier}, {Johansen}, {Janson}  \&
  {Rafikov}}{{Chabrier} et~al.}{2014}]{chabrier2014}
{Chabrier} G.,  {Johansen} A.,  {Janson} M.,   {Rafikov} R.,  2014, \mn@doi
  [Protostars and Planets VI] {10.2458/azu_uapress_9780816531240-ch027}, \href
  {http://adsabs.harvard.edu/abs/2014prpl.conf..619C} {pp 619--642}

\bibitem[\protect\citeauthoryear{{Correia, A. C. M.}, {Udry, S.}, {Mayor, M.},
  {kar, J. La}, {Naef, D.}, {Pepe, F.}, {Queloz, D.}  \& {Santos, N.
  C.}}{{Correia, A. C. M.} et~al.}{2005}]{correia2005}
{Correia, A. C. M.} {Udry, S.} {Mayor, M.} {kar, J. La} {Naef, D.} {Pepe, F.}
  {Queloz, D.}  {Santos, N. C.} 2005, \mn@doi [A&A]
  {10.1051/0004-6361:20042376}, 440, 751

\bibitem[\protect\citeauthoryear{{Degroote, P.}, {Conroy, K.}, {Hambleton, K.},
  {Bloemen, S.}, {Pablo, H.}, {Giammarco, J.}  \& {Prsa, A.}}{{Degroote, P.}
  et~al.}{2013}]{degroote2013}
{Degroote, P.} {Conroy, K.} {Hambleton, K.} {Bloemen, S.} {Pablo, H.}
  {Giammarco, J.}  {Prsa, A.} 2013, \mn@doi [EAS Publ. Series]
  {10.1051/eas/1364038}, 64, 277

\bibitem[\protect\citeauthoryear{Doyle et~al.,}{Doyle et~al.}{2011}]{doyle2011}
Doyle L.~R.,  et~al., 2011, \mn@doi [Science] {10.1126/science.1210923}, 333,
  1602

\bibitem[\protect\citeauthoryear{{Fabrycky}}{{Fabrycky}}{2010}]{fabrycky2010}
{Fabrycky} D.~C.,  2010, preprint, \href
  {http://adsabs.harvard.edu/abs/2010arXiv1006.3834F} {} (\mn@eprint {arXiv}
  {1006.3834})

\bibitem[\protect\citeauthoryear{Gazak, Johnson, Tonry, Dragomir, Eastman, Mann
   \& Agol}{Gazak et~al.}{2012}]{gazak2012}
Gazak J.~Z.,  Johnson J.~A.,  Tonry J.,  Dragomir D.,  Eastman J.,  Mann A.~W.,
    Agol E.,  2012, \mn@doi [Advances in Astron.] {10.1155/2012/697967}, 2012

\bibitem[\protect\citeauthoryear{Hinse, Horner  \& Wittenmyer}{Hinse
  et~al.}{2014}]{hinse2014}
Hinse T.~C.,  Horner J.,   Wittenmyer R.~A.,  2014, \mn@doi [JASS]
  {10.5140/JASS.2014.31.3.187}, 31, 187

\bibitem[\protect\citeauthoryear{Kane \& von Braun}{Kane \& von
  Braun}{2008}]{kane2008}
Kane S.~R.,  von Braun K.,  2008, \mn@doi [Proc. IAU]
  {10.1017/S1743921308026641}, 4, 358

\bibitem[\protect\citeauthoryear{Kostov et~al.,}{Kostov
  et~al.}{2016}]{kostov2016}
Kostov V.~B.,  et~al., 2016, \mn@doi [ApJ] {10.3847/0004-637X/827/1/86}, 827,
  86

\bibitem[\protect\citeauthoryear{Lee, Kim, Kim, Koch, Lee, Kim  \& Park}{Lee
  et~al.}{2009}]{lee2009}
Lee J.~W.,  Kim S.-L.,  Kim C.-H.,  Koch R.~H.,  Lee C.-U.,  Kim H.-I.,   Park
  J.-H.,  2009, \mn@doi [AJ] {10.1088/0004-6256/137/2/3181}, 137, 3181

\bibitem[\protect\citeauthoryear{Mandel \& Agol}{Mandel \&
  Agol}{2002}]{mandel2002}
Mandel K.,  Agol E.,  2002, \mn@doi [ApJ] {10.1086/345520}, 580, L171

\bibitem[\protect\citeauthoryear{Meschiari \& Laughlin}{Meschiari \&
  Laughlin}{2010}]{meschiari2010}
Meschiari S.,  Laughlin G.~P.,  2010, \mn@doi [ApJ]
  {10.1088/0004-637X/718/1/543}, 718, 543

\bibitem[\protect\citeauthoryear{Meschiari, Wolf, Rivera, Laughlin, Vogt  \&
  Butler}{Meschiari et~al.}{2009}]{meschiari2009}
Meschiari S.,  Wolf A.~S.,  Rivera E.,  Laughlin G.,  Vogt S.,   Butler P.,
  2009, \mn@doi [PASP] {10.1086/605730}, 121, 1016

\bibitem[\protect\citeauthoryear{Nordlund}{Nordlund}{2011}]{nordlund2011}
Nordlund A.,  2011, \mn@doi [Proc. IAU] {10.1017/S1743921311020023}, 6, 105

\bibitem[\protect\citeauthoryear{Orosz et~al.,}{Orosz et~al.}{2012}]{orosz2012}
Orosz J.~A.,  et~al., 2012, \mn@doi [Science] {10.1126/science.1228380}, 337,
  1511

\bibitem[\protect\citeauthoryear{Pecaut \& Mamajek}{Pecaut \&
  Mamajek}{2013}]{pecaut2013}
Pecaut M.~J.,  Mamajek E.~E.,  2013, \mn@doi [ApJS]
  {10.1088/0067-0049/208/1/9}, 208, 9

\bibitem[\protect\citeauthoryear{Prsa \& Zwitter}{Prsa \&
  Zwitter}{2005}]{prsa2005}
Prsa A.,  Zwitter T.,  2005, \mn@doi [ApJ] {10.1086/430591}, 628, 426

\bibitem[\protect\citeauthoryear{Prsa et~al.,}{Prsa et~al.}{2011}]{prsa2011}
Prsa A.,  et~al., 2011, \mn@doi [AJ] {10.1088/0004-6256/141/3/83}, 141, 83

\bibitem[\protect\citeauthoryear{Sigurdsson, Richer, Hansen, Stairs  \&
  Thorsett}{Sigurdsson et~al.}{2003}]{sigurdsson2003}
Sigurdsson S.,  Richer H.~B.,  Hansen B.~M.,  Stairs I.~H.,   Thorsett S.~E.,
  2003, \mn@doi [Science] {10.1126/science.1086326}, 301, 193

\bibitem[\protect\citeauthoryear{Slawson et~al.,}{Slawson
  et~al.}{2011}]{slawson2011}
Slawson R.~W.,  et~al., 2011, \mn@doi [AJ] {10.1088/0004-6256/142/5/160}, 142,
  160

\bibitem[\protect\citeauthoryear{{Southworth}, {Maxted}  \&
  {Smalley}}{{Southworth} et~al.}{2004}]{southworth2004}
{Southworth} J.,  {Maxted} P.~F.~L.,   {Smalley} B.,  2004, \mn@doi [\mnras]
  {10.1111/j.1365-2966.2004.07871.x}, \href
  {http://adsabs.harvard.edu/abs/2004MNRAS.351.1277S} {351, 1277}

\makeatother
\end{thebibliography}

%%%%%%%%%%%%%%%%%%%%%%%%%%%%%%%%%%%%%%%%%%%%%%%%%%

% Don't change these lines
\bsp	% typesetting comment
\label{lastpage}
\end{document}